\RequirePackage{ifpdf}
\ifpdf 
\documentclass[pdftex]{sigma}
\else
\documentclass{sigma}
\fi

\begin{document}

\allowdisplaybreaks

\renewcommand{\PaperNumber}{066}

\FirstPageHeading

\ShortArticleName{Tools for Verifying Classical and Quantum Superintegrability}

\ArticleName{Tools for Verifying Classical\\ and Quantum Superintegrability}

\Author{Ernest  G.~KALNINS~$^\dag$, Jonathan M.~KRESS~$^\ddag$ and Willard MILLER Jr.$^\S$}

\AuthorNameForHeading{E.G.~Kalnins, J.M.~Kress and W.~Miller Jr.}

\Address{$^\dag$~Department of Mathematics, University of Waikato, Hamilton, New Zealand}
\EmailD{\href{mailto:math0236@math.waikato.ac.nz}{math0236@math.waikato.ac.nz}}
\URLaddressD{\url{http://www.math.waikato.ac.nz}}

\Address{$^\ddag$~School of Mathematics, The University of New South Wales,
Sydney NSW 2052, Australia}
\EmailD{\href{mailto:j.kress@unsw.edu.au}{j.kress@unsw.edu.au}}
\URLaddressD{\url{http://web.maths.unsw.edu.au/~jonathan/}}

\Address{$^\S$~School of Mathematics, University of Minnesota,
 Minneapolis, Minnesota,55455, USA}
\EmailD{\href{mailto:miller@ima.umn.edu}{miller@ima.umn.edu}}
\URLaddressD{\url{http://www.ima.umn.edu/~miller/}}

\ArticleDates{Received June 04, 2010, in f\/inal form August 06, 2010;  Published online August 18, 2010}

\Abstract{Recently  many new classes of integrable systems in $n$ dimensions occurring in
classical  and quantum mechanics have been shown to  admit a functionally independent set of $2n-1$
symmetries  polynomial in the canonical momenta, so that they are in fact superintegrable. These newly discovered systems are all separable in some coordinate system and, typically, they depend on one or more parameters in such a way that the system is superintegrable exactly when some of  the parameters are rational numbers. Most of the constructions to date are for $n=2$ but cases where $n>2$ are multiplying rapidly. In
this article we organize a large class of such systems, many new,  and emphasize the
underlying mechanisms which enable this phenomena to  occur and to prove superintegrability. In addition to proofs of classical superintegrability we show that the 2D caged anisotropic oscillator and a~St\"ackel transformed version on the 2-sheet hyperboloid are quantum superintegrable for all rational relative frequencies,  and that a  deformed 2D Kepler--Coulomb system is  quantum superintegrable for all rational values of a parameter $k$ in the potential.}


\Classification{20C99; 20C35; 22E70}

\section{Introduction}
We def\/ine  an $n$-dimensional classical superintegrable system to be  an integrable Hamiltonian
system that not only possesses $n$ mutually Poisson -- commuting constants of the motion, but in addition,
the Hamiltonian Poisson-commutes with $2n-1$ functions on the phase space that are globally
def\/ined and polynomial in the momenta. Similarly, we def\/ine a quantum
superintegrable system to be a quantum Hamiltonian which is one of a set
of $n$ independent mutually commuting dif\/ferential operators, and that commutes with a set of $2n-1$
independent dif\/ferential opera\-tors of f\/inite order. We restrict to classical systems of the form ${\cal H}=\sum_{i,j=1}^ng^{ij}p_ip_j+V$ and corresponding quantum systems $H=\Delta_n+{\tilde V}$. These systems, including the classical Kep\-ler and anisotropic oscillator problems and the quantum anisotropic oscillator and hydrogen atom have great historical importance, due to their remarkable properties,~\cite{SCQS,IMA}. The order of a classical superintegrable system is the maximum order of the generating constants of the motion (with the Hamiltonian excluded) as a polynomial in the momenta, and the maximum order of the quantum symmetries as dif\/ferential operators.  Systems of 2nd order have been well studied and there is now a structure and classif\/ication theory~\cite{ KKM20061,KKMP, DASK2005, KKM2007,KMP2007,KMP2008}.  For 3rd and higher order superintegrable systems much less is known. In particular there have been  relatively few examples and there  is almost no structure theory. However, within the last three years there has been a dramatic increase in discovery of new families of possible higher order superintegrable classical and quantum systems~\cite{CDR2008, RTW, Evans2008a,TTW,TTW2,CQ10,TW2010, PW2010}. The authors and collaborators have developed methods for verifying superintegrability of these proposed systems,~\cite{KMP10,KKM10,KKM10a,KMPog10}. In the cited papers the emphasis was on particular systems of special importance and recent interest. Here, however, the emphasis is on  the methods themselves.

In Section~\ref{section2} we review a method for constructing classical constants of the motion of all orders for $n$-dimensional Hamiltonians that admit a
separation of variables in some orthogonal coordinate system. Then we apply it to the case $n=2$ to show how to derive superintegrable systems for all values of a rational parameter $k$ in the potential. Many of our examples are new. In Section~\ref{section3} we review our method for establishing a canonical form  for quantum symmetry operators of all orders for $2$-dimensional Schr\"odinger operators such that  the Schr\"odinger eigenvalue equation  admits a
separation of variables in some orthogonal coordinate system. Then we apply this method to establish the quantum superintegrability of the caged anisotropic oscillator for all frequencies that are rationally related, and for a St\"ackel transformed version of this system on the 2-sheet hyperboloid. We  give a second proof for the caged oscillator in all dimensions~$n$ that relies on recurrence relations for associated Laguerre polynomials. We also  apply the canonical equations to establish the quantum superintegrability of a deformed Kepler--Coulomb system for all rational values of a parameter~$k$ in the potential.

\subsection{The construction tool for classical systems}  \label{subsection1}

There are far  more verif\/ied superintegrable   Hamiltonian systems in
classical mechanics than was the case  3 years ago. The principal method for constructing  and verifying these new systems requires that the system is already integrable, in particular, that it admits a~separation of variables in some coordinate system. For a Hamiltonian system in $2n$-dimensional phase space the separation gives us $n$ second order constants of the motion in involution. In this paper we f\/irst review a  general procedure, essentially the construction of action angle variables,  which
 yields an additional $n-1$ constants, such that the set of $2n-1$ is functionally independent. This is of little interest unless it is possible to extract $n$ new constants of the motion that are polynomial in the momenta, so that the system is superintegrable.  We will show how this can be done in many cases. We start f\/irst with the  construction of action angle variables for Hamiltonian systems in $n$ dimensions, and later specialize to the case $n=2$ to verify superintegrability.

Consider a classical system in $n$ variables on a complex Riemannian manifold that admits separation of variables in orthogonal separable coordinates $x_1,\dots, x_n$.
Then there is an $n\times n$ St\"ackel matrix
\[ S=\left( S_{ij}(x_i)\right)\]
 such that $\Phi=\det S\ne0$ and  the Hamiltonian is
\[ {\cal H}={\cal L}_1=\sum _{i=1}^n T_{1i}\left(p_i^2+v_i(x_i)\right)=\sum _{i=1}^n T_{1i}\ p_i^2+V(x_1,\dots, x_n) ,\]
where $V=\sum _{i=1}^n T_{1i}\ v_i(x_i)$, $T$ is the matrix inverse to $S$:
\begin{gather} \label{inversematrix}
\sum _{j=1}^nT_{ij}S_{jk}=\sum _{j=1}^n S_{ij}T_{jk}=\delta_{ik}, \qquad 1\le i,k\le n,
\end{gather}
and $\delta_{ik}$ is the Kronecker delta. Here, we must require $ \Pi_{i=1}^n T_{1i}\ne 0$. We def\/ine the quadratic constants of the motion ${\cal L}_k$, $k=1,\dots, n$ by
\begin{gather*}
{\cal L}_k=\sum_{i=1}^nT_{ki}\big(p_i^2+v_i\big),\qquad k=1,\dots, n,
\end{gather*}
or
\begin{gather}\label{varsep} p_i^2+v_i=\sum_{j=1}^n S_{ij}{\cal L}_j,\qquad 1\le i\le n.\end{gather}
As is well known,
\[ \{{\cal L}_j,{\cal L}_k\}=0,\qquad 1\le j,k\le n.
\]
Here,
\[ \{{\cal A}({\bf x}, {\bf p}), {\cal B}({\bf x}, {\bf p})\}=\sum_{i=1}^n(\partial_i {\cal A}\ \partial_{p_i}{\cal B}-\partial_{p_i}{\cal A}\ \partial_i {\cal  B}).\]
Furthermore, by dif\/ferentiating identity (\ref{inversematrix}) with respect to $x_h$  we obtain
\begin{gather*}
\partial_h T_{i\ell}=-\sum_{j=1}^n T_{ih}S'_{hj}T_{j\ell},\qquad 1\le h,i,\ell\le n,
\end{gather*}
where $S'_{hj}=\partial_hS_{hj}$.

Now we def\/ine nonzero functions $M_{kj}(x_j,p_j, {\cal L}_1,\dots,{\cal L}_n)$ on the manifold by the requirement
\[ \{M _{kj},{\cal L}_\ell\}=T_{\ell j}S_{jk},\qquad 1\le k,j,\ell\le n.\] It is straightforward to check that these conditions are equivalent to the dif\/ferential equations
\begin{gather*}
 2p_j\partial_j M_{kj}+\Bigg(-v'_j+\sum_{q=1}^n S'_{jq}{\cal L}_q\Bigg)\partial_{p_j}M_{kj}=S_{jk},\qquad 1\le j,k\le n.\end{gather*}
We can use the equalities (\ref{varsep}) to  consider $M_{kj}$ either as a function of $x_j$ alone, so that $\frac{d}{dx_j}M_{kj}=S_{jk}/2p_j$ or as a function of $p_j$ alone. In this paper we will take the former point of view.

Now def\/ine functions
\begin{gather*}
{\tilde {\cal L}}_q=\sum_{j=1}^n M_{qj},\qquad 1\le q\le n.
\end{gather*}
Then we have
\begin{gather}\label{cancomm} \{{\tilde{\cal  L}}_q,{\cal L}_\ell\}=\sum_{j=1}^n T_{\ell j}S_{jq}=\delta_{\ell q}.\end{gather}
This shows that the $2n-1$ functions
\[{\cal H}={\cal L}_1, {\cal L}_2,\dots,{\cal L}_n, {\tilde {\cal L}}_2,\dots, {\tilde{\cal L}}_n,\]
are constants of the motion and, due to relations~(\ref{cancomm}), they are functionally independent.

Now let's  consider how this construction works in $n=2$ dimensions.  By replacing each separable coordinate by a suitable function of itself and the constants of the motion by suitable linear combinations of themselves, if necessary, e.g. \cite{EIS},  we can always assume that  the St\"ackel matrix and its inverse are  of the form
\begin{gather*}
S=\left(\begin{array}{ll} f_1&1\\ f_2&-1\end{array}\right),\qquad T=\frac{1}{f_1+f_2}\left(\begin{array}{ll} 1&1\\ f_2 &-f_1\end{array}\right),\end{gather*}
where $f_j$ is a function of the variable $x_j$ alone. The constants of the motion ${\cal L}_1={\cal H}$ and ${\cal L}_2$ are given to us via variable separation. We want to compute a new constant of the motion ${\tilde {\cal L}}_2$ functionally independent of ${\cal L}_1$, ${\cal L}_2$.  Setting $M_{21}=M,\ M_{22}=-N$, we see that
\begin{gather}\label{MN1} 2p_1\frac{d}{dx_1}M=1,\qquad 2p_2\frac{d}{dx_2}N=1,\end{gather}
from which we can determine~$M$,~$N$. Then ${\tilde {\cal L}}_2=M-N$ is the constant of the motion that we seek.

  The treatment  of subgroup separable superintegrable systems in $n$ dimensions,~\cite{KKM10}, is also a~special case of the above construction. Suppose the St\"ackel matrix takes the form
\begin{gather*}
S=\left(\begin{array}{cccccc} 1&-f_1,&0& \cdots &\cdots&0\\ 0&1&-f_2 &0&\cdots&0\\ 0&0&\ddots& \ddots&\ddots& 0\\ \hdots&\hdots&\hdots&\hdots&\hdots&\hdots\\0&0&0&\cdots&1&-f_{n-1}\\ 0&0&0&\cdots&0&1\end{array}\right)
\end{gather*}
where $f_i=f_i(x_i)$, $i=1,\dots,n$. The inverse matrix in this case is
\begin{gather*}
T=\left(\begin{array}{cccccc} 1&f_1& f_1f_2& \cdots &\cdots&f_1f_2\cdots  f_{n-1}\\ 0&1& f_2 &f_2f_3&\cdots&f_2 \cdots f_{n-1}\\
0&0&\ddots&\ddots&\ddots&\vdots\\\hdots&\hdots&\hdots&\hdots&\hdots&\hdots\\ 0&0&0&\cdots&1&f_{n-1}\\0&0&0&\cdots&0&1\end{array}\right),\end{gather*}
which leads exactly to the construction found in~\cite{KKM10}.

\subsection[Application of the construction for $n=2$]{Application of the construction for $\boldsymbol{n=2}$}

As we have seen, for $n=2$ and separable coordinates $u_1=x$, $u_2=y$ we have
\begin{gather*} 
{\cal H}={\cal L}_1=\frac{1}{f_1(x)+f_2(y)}\big(p_x^2+p_y^2+v_1(x)+v_2(y)\big),\\
  {\cal  L}_2 =\frac{f_2(y)}{f_1(x)+f_2(y)}\left(p_x^2+v_1(x)\right)-\frac{f_1(x)}{f_1(x)+f_2(y)}\left(p_y^2+v_2(y)\right).
  \nonumber
\end{gather*}
We will present strategies for determining functions $f_1$, $f_2$, $v_1$, $v_1$ such that there exists a 3rd constant of the motion, polynomial in the momenta.
The constant of the motion ${\tilde {\cal L}}_2=M-N$  constructed by solving equations~(\ref{MN1}) is usually not a polynomial in the momenta, hence not directly useful in verifying superintegrability. We describe a procedure for obtaining a polynomial constant from $M-N$, based on the observation that the integrals
\[M=\frac12\int \frac{dx_1}{\sqrt{f_1{\cal H}+{\cal L}_2-v_1}},\qquad N=\frac12\int \frac{dx_2}{\sqrt{f_2{\cal H}-{\cal L}_2-v_2}},\]
can often be expressed in terms of multiples of the inverse hyperbolic sine or cosine (or the ordinary inverse sine or cosine), and the hyperbolic sine and cosine satisfy addition formulas. Thus we will search for functions $f_j$, $v_j$ such that $M$ and $N$ possess this property. There is a~larger class of prototypes for this construction, namely the second order superintegrable systems. These have already been classif\/ied for 2-dimensional constant curvature spaces~\cite{KKMP} and, due to the fact that every superintegrable system on a Riemannian or pseudo-Riemannian space in two dimensions is St\"ackel equivalent to a constant curvature superintegrable system \cite{KKM2,fine}, this list includes all cases. We will typically start our construction with one of these second order systems and add parameters to get a family of higher order superintegrable systems.

A basic observation is that to get inverse trig functions for the integrals $M$, $N$  we can choose the potential functions $f(z)$, $v(z)$ from the list
\begin{gather*}
Z_1(z)=Az^2+Bz+C,\qquad Z_2(z)= \frac{A+B\sin pz}{ \cos ^2py},\\
Z_3(z)= \frac{A}{ \cos ^2pz} + \frac{B}{ \sin ^2pz},\qquad Z_4(z)=Ae^{2ipz}+Be^{ipz},\nonumber\\
Z_5(z)=Az^2+\frac{B}{ z^2},\qquad Z_6(z)=\frac{A}{ z} + \frac{B}{ z^2},\nonumber
\end{gather*}
sometimes restricting the parameters to special cases.
Many of these cases actually occur in the lists~\cite{KKMP} so we are guaranteed that it will be possible to construct at least one second order polynomial constant of the motion from such a selection. However, some of the cases lead only to higher order constants.

\subsection*{Cartesian type systems}

For simplicity, we begin with Cartesian coordinates in f\/lat space. The systems of this type are related to oscillators and are associated with functions $Z_j$ for $j=1,5,6$. We give two examples.

\medskip\noindent {\bf [E1]}.
For our f\/irst construction we give yet another verif\/ication that the extended caged harmonic oscillator is classically superintegrable.
We  modify the second order superintegrable system [E1], \cite{KKMP},  by looking at the potential
\[ V=\omega ^2_1x^2+\omega ^2_2y^2+ \frac{\beta }{ x^2} +\frac{\gamma }{ y^2},\]
with ${\cal L}_2=p^2_x+\omega ^2_1x^2+ \frac{\beta }{ x^2}$. (For $\omega_1=\omega_2$ this is just [E1].) It corresponds to the choice of a~function of the form $Z_5$ for each of $v_1$, $v_2$.
Evaluating the integrals,   we obtain the solutions
\begin{gather*}
 M(x,p_x)= \frac{i}{ 4\omega _1} {\cal A},\qquad \sinh{\cal A }= \frac{i(2\omega ^2_1x^2-{\cal L}_2)}{ \sqrt {{\cal L}^2_2-4\omega ^2_1\beta }},\\
 N(y,p_y)=\frac{-i}{ 4\omega _2}{\cal  B},\qquad \sinh{\cal  B }= \frac{i(2\omega ^2_2y^2-{\cal H}+{\cal L}_2)}{ \sqrt {({\cal H}-{\cal L}_2)^2-4\omega ^2_2\gamma }},
 \end{gather*}
to within arbitrary additive constants. We choose these constants so that $M$, $N$ are proportional to inverse hyperbolic sines. Then, due to the formula $\cosh^2 u-\sinh^2 u=1$, we can use the identities (\ref{varsep}) to compute $\cosh {\cal  A}$ and $\cosh {\cal B}$:
\[
\cosh{\cal  A }= \frac{2\omega _1xp_x}{ \sqrt {{\cal L}^2_2-4\omega ^2_1\beta }},\qquad \cosh{\cal B }= \frac{2\omega _2yp_y}{ \sqrt{({\cal H}-{\cal L}_2)^2-4\omega ^2_2\gamma }}.\]
Now suppose that  $\omega _1/\omega _2=k$ is rational, i.e.\ $k=\frac{p}{q}$ where $p$, $q$ are relatively prime integers. Then $\omega_1=p\omega$, $\omega_2=q\omega$ and
\[\sinh(-4ipq\omega[M-N])=\sinh(q{\cal A}+p{\cal B}),\qquad \cosh(-4ipq\omega[M-N])=\cosh(q{\cal A}+p{\cal B}),\]
are both constants of the motion. Each of these will lead to a polynomial constant of the motion. Indeed, we can use the relations
\begin{gather*}
 (\cosh x\pm\sinh x)^n=\cosh nx\pm\sinh nx,\\
   \cosh (x+y)=\cosh x\cosh y+\sinh x\sinh y,\\
   \sinh (x+y)=\cosh x\sinh y+\sinh x\cosh y,\\
 \cosh nx=\sum_{j=0}^{[n/2]} \left( \begin{array}{c}n\\ 2j\end{array}\right) \sinh ^{2j}x\ \cosh^{n-2j}x,\\
  \sinh nx=\sinh x\sum_{j=1}^{[(n=1)/2]} \left( \begin{array}{c}n\\ 2j-1\end{array}\right) \sinh ^{2j-2}x  \cosh^{n-2j-1}x.
\end{gather*}
recursively to express each constant as a polynomial in $\cosh {\cal A}$, $\sinh{\cal A}$, $\cosh {\cal B}$, $\sinh{\cal B}$.  Then, writing each constant as a single fraction with denominator of the form
\[  (({\cal H}-{\cal L}_2)^2-4\omega ^2_2\gamma )^{n_1 /2}({{\cal L}^2_2-4\omega ^2_1\beta })^{n_2 /2}\]
we see that the numerator is a polynomial constant of the motion. Note that, by construction, both $\sinh(-4ipq\omega[M-N])$ and $ \cosh(-4ipq\omega[M-N])$ will have nonzero Poisson brackets with ${\cal L}_2$, hence they are each functionally independent of ${\cal H}$, ${\cal L}_2$. Since each of our polynomial constants of the motion dif\/fers from these by a factor that is a function of ${\cal H}$, ${\cal L}_2$ alone, each polynomial constant of the motion is also functionally independent of ${\cal H}$, ${\cal L}_2$. Similar remarks apply to all of our examples.

\medskip\noindent {\bf [E2]}. There are several proofs of superintegrability for this system, but we add another.
Here, we make the choice $v_1=Z_1$, $v_2=Z_5$, corresponding to the second order superintegrable system~[E2] in~\cite{KKMP}.  The potential is
\[ V=\omega ^2_1x^2+\omega ^2_2y^2+\alpha x + \frac{\beta }{ y^2},\]
where ${\cal L}_2=p^2_x+\omega ^2_1x^2+\alpha x$
and the system is second order superintegrable for the case $\omega_1=\omega_1$.
Applying our method we obtain
\begin{gather*}
 M(x,p_x)=\frac{i}{ 2\omega _1} {\cal A},\qquad \sinh{\cal A}= \frac{i(\omega ^2_1x+\alpha )}{ \sqrt {4\omega ^2_1{\cal L}_2+\alpha ^2}},\qquad
\cosh{\cal A}=\frac{2\omega _1p_x}{ \sqrt{4\omega ^2_1{\cal L}_2+\alpha ^2}},\\
 N(y,p_y)=\frac{i}{ 4\omega _2} {\cal B},\qquad
\sinh{\cal B}= \frac{i(2\omega ^2_2y^2-{\cal H}+{\cal L}_2)}{ \sqrt {({\cal H}-{\cal L}_2)^2-4\omega ^2_2\beta }},\qquad
\cosh{\cal  B } = \frac{2\omega _2yp_y}{ \sqrt{({\cal H}-{\cal L}_2)^2-4\omega ^2_2\beta }}.
\end{gather*}
Thus  if $\omega _1/2\omega _2$ is rational we obtain a
constant of the motion which is polynomial in the momenta.

\subsection*{Polar type systems}

Next we look at f\/lat space systems that separate in polar coordinates. The Hamiltonian is of the form
\begin{gather*}
  {\cal H}=p_r^2+\frac{1}{r^2}\big(p_\theta^2+f(r)+g(\theta)\big)=e^{-2R}\big(p^2_R+p^2_\theta +v_1(R)+v_2(\theta)\big),\\
 x=R=\ln r,\qquad y=\theta,\qquad f_1=e^{2R},\qquad f_2=0.
\end{gather*}
 Cases for which the whole process works can now be evaluated. Possible choices of $f$ and $g$ are
\begin{alignat*}{5}
 & (1) \ &&  f(r)=\alpha r^2,\quad g(\theta )=Z_3(k\theta ),\qquad && (2) \ &&  f(r)=\frac{\alpha }{ r} ,\quad g(\theta )=Z_3(k\theta ),& \\
& (3) \ && f(r)=\alpha r^2 ,\quad g(\theta )=Z_4(k\theta ),\qquad && (4) \ &&  f(r)=\frac{\alpha }{ r} ,\quad  g(\theta )=Z_4(k\theta ), & \\
 & (5) \ &&  f(r)=\alpha r^2 ,\quad g(\theta )=Z_2(k\theta ),\qquad && (6) \ && f(r)=\frac{\alpha }{ r} ,\quad g(\theta )=Z_2(k\theta ). &
 \end{alignat*}
In each case if $p$ is rational there is an extra constant of the motion that
is polynomial in the canonical momenta.

\medskip \noindent {\bf [E1]}.
{\bf Case (1)} is system [E1] for $k=1$. For general $k$ this is the TTW system \cite{TTW}, which we have shown to be supperintegrable for $k$ rational.

\medskip \noindent {\bf Case (2).}
This case is not quadratic superintegrable, but as shown in \cite{KKM10a} it is superintegrable for all rational $k$.

\medskip\noindent {\bf [E8]}.
For our next example we take {\bf Case (3)} where $z=x+iy$:
\[ V=\alpha z\bar z +\beta  \frac{z^{k-1}}{ {\bar z}^{k+1}} +\gamma \frac {z^{{k /2}-1}}{{\bar z}^{k/2+1}}.\]
for arbitrary $k$. (If $k=2$ this is the nondegenerate superintegrable system [E8] listed in~\cite{KKMP}.) In polar coordinates, with variables $r=e^R$ and
$z=e^{R+i\theta }$. Then we have
\begin{gather*}
{\cal H}=e^{-2R}\big(p^2_R+p^2_\theta +4\alpha e^{4R}+\beta e^{2ik\theta }+\gamma e^{ik\theta }\big),\\
 -{\cal L}_2=p^2_\theta +\beta e^{2ik\theta }+\gamma e^{ik\theta },\qquad {\cal H}=e^{-2R}\big(p^2_R-{\cal L}_2+4\alpha e^{4R}\big).\nonumber
\end{gather*}
 Our method yields
\begin{gather*}
N(\theta ,p_\theta )=\frac{i}{ k\sqrt{ -{\cal L}_2}}{\cal  B},\qquad \sinh{\cal B} =
\frac{i(2{\cal L}_2e^{-ik\theta }+\gamma )}{ \sqrt {-4\beta {\cal L}_2+\gamma ^2}},\qquad
\cosh{\cal  B}= \frac{2ip_\theta \sqrt{-{\cal L}_2}}{\sqrt {-4\beta {\cal L}_2+\gamma ^2}}
e^{-ik\theta },\\
 M(R,p_R)= \frac{i}{ \sqrt{- {\cal L}_2}}{\cal A},\qquad \sinh{\cal A} =\frac{(-2{\cal L}_2e^{-2R}-H)}{\sqrt{-4\alpha {\cal L}_2-{\cal H}^2}},\qquad
\cosh{\cal  A}= \frac{2ip_R\sqrt{-{\cal L}_2}}{ \sqrt {-4\alpha {\cal L}_2-{\cal H}^2}} e^{-2R}.
\end{gather*}
This system is superintegrable for all rational $k$.

\medskip\noindent {\bf [E17]}.
Taking {\bf Case (4)} we have, for $z=x+iy$:
\[ V = \frac{\alpha}{\sqrt{ z\bar z}} +\beta \frac{ {\bar z}^{k-1}}{ z^{k+1}}
+\gamma  \frac{{\bar z}^{k/2-1}}{z^{k/2+1}}.\]
(For $k=1$ this is the superintegrable system [E17] in \cite{KKMP}.) Then we have
\begin{gather}\label{hamE17}H=e^{-2R}\big(p^2_R+p^2_\theta +\alpha e^R+\beta e^{-2ik\theta }+\gamma e^{-ik\theta }\big),\\
 -{\cal L}_2=p^2_\theta +\beta e^{-2ik\theta }+\gamma e^{-ik\theta },\qquad
{\cal H}=e^{-2R}\big(p^2_R-{\cal L}_2+\alpha e^R\big).\nonumber
\end{gather}
Applying our procedure we f\/ind the functions
\begin{gather*}
N(\theta ,p_\theta )=\frac{1}{ k\sqrt{ {\cal L}_2}} {\cal B},\qquad\sinh{\cal B }=
\frac{(2{\cal L}_2e^{ik\theta }+\gamma )}{\sqrt{4\beta {\cal L}_2-\gamma ^2}},\qquad
\cosh{\cal B }=- \frac{2ip_\theta \sqrt{{\cal L}_2}}{ \sqrt {4\beta {\cal L}_2-\gamma ^2}}
e^{ik\theta },\\
 M(R,p_R)=\frac{1}{ \sqrt{ {\cal L}_2}}{\cal A},\qquad\sinh{\cal A} =\frac{(-2{\cal L}_2e^{-R}+\alpha )}{ \sqrt {4{\cal HL}_2-\alpha ^2}},\qquad
\cosh{\cal  A }= \frac{2\sqrt{{\cal L}_2}p_R}{ \sqrt {4{\cal HL}_2-\alpha ^2}} e^{-R}.
\end{gather*}
This demonstrates superintegrability for all rational~$k$.

\medskip\noindent {\bf [E16]}.
{\bf Case (5)} corresponds to [E16] for $k=1$, and our method shows that it is superintegrable for all rational $k$.

\medskip \noindent {\bf Case (6).}
This case is not quadratic superintegrable, but  it is superintegrable for all rational~$k$.

\subsection*{Spherical type  systems}

These are systems that separate   in spherical type coordinates on the complex 2-sphere.  The Hamiltonian is of the form
\begin{gather*}
{\cal H}=\cosh^2\psi\big(p_\psi^2+p_\varphi^2+v_1(\varphi)+v_2(\psi)\big),\\
 x=\varphi,\qquad y=\psi,\qquad f_1=0,\qquad f_2=\frac{1}{\cosh^2\psi}.
\end{gather*}
 Embedded in complex Euclidean 3-space with Cartesian coordinates, such 2-sphere systems can be written in the form
\[  H={\cal J}_1^2+{\cal J}_2^2+{\cal J}_3^2 +V({\bf s}),\]
where
\begin{gather*}
 {\cal J}_3=s_1p_{s_2}-s_2p_{s_1},\qquad {\cal J}_1=s_2p_{s_3}-s_3p_{s_2},\qquad {\cal J}_2=s_3p_{s_1}-s_1p_{s_3},\qquad  s_1^2+s_2^2+s_3^3=1.
 \end{gather*}

\medskip\noindent {\bf [S9]}. Here, we have the case $v_1(\varphi)=Z_3$ and $v_2(\psi) $ is a special case of $Z_3$.
\[ V = \frac{\alpha }{ s^2_1} + \frac{\beta }{ s^2_2} + \frac{\gamma }{ s^2_3}.\]
It is convenient to choose spherical coordinates
\[ s_1 = \frac{\cos\varphi }{ \sinh\psi },\qquad s_2 =  {\sin\varphi }\ { \cosh\psi },\qquad
s_3=\tanh\psi.\]
In terms of these coordinates the Hamiltonian has the form
\begin{gather*}
 {\cal H}=\cosh^2\psi \left[p^2_\psi +p^2_\varphi + \frac{\alpha }{ \cos ^2\varphi } +
\frac{\beta }{ \sin ^2\varphi } + \frac{\gamma }{\sinh^2\psi } \right],\\
  {\cal L}_2=p^2_\varphi + \frac{\alpha }{ \cos ^2\varphi } +
\frac{\beta }{ \sin ^2\varphi },\qquad  H=\cosh^2\psi \left[p^2_\psi +{\cal L}_2+\frac{\gamma }{ \sinh^2\psi } \right],
\end{gather*}
system [S9] in \cite{KKMP}.
We extend this Hamiltonian via  the replacement $\varphi \rightarrow k\varphi $ and proceed
with our method. Thus
\begin{gather*} 
 {\cal H}=\cosh^2\psi \left[p^2_\psi +p^2_\varphi + \frac{\alpha }{\cos ^2k\varphi } +
\frac{\beta }{\sin ^2k\varphi } + \frac{\gamma }{ \sinh^2\psi } \right],\\
  {\cal L}_2=p^2_\varphi + \frac{\alpha }{ \cos ^2k\varphi } + \frac{\beta }{ \sin ^2k\varphi }
  \nonumber
\end{gather*}
with $\cal H$ expressed as above. The functions that determine the extra constant are
\begin{gather*}
M(\varphi ,p_\varphi )= \frac{i}{4k\sqrt{ {\cal L}_2}} {\cal A},\\ \sinh{\cal  A }=
\frac{i({\cal L}_2\cos (2k\varphi )-\alpha +\beta )}{ \sqrt {({\cal L}_2-\alpha -\beta )^2-4\alpha \beta }},\qquad
\cosh{\cal A}=\frac{\sin (2k\varphi )p_\varphi }{ \sqrt {({\cal L}_2-\alpha -\beta )^2-4\alpha \beta }},\\
 N(\psi ,p_\psi )= \frac{i}{ 4\sqrt {{\cal L}_2}}{\cal B},\\ \sinh{\cal B }=
\frac{i({\cal L}_2\cosh(2\psi )+\gamma -{\cal H})}{ \sqrt {({\cal H}+{\cal L}_2-\gamma )^2+4{\cal L}_2\gamma }},\qquad \cosh{\cal B }= \frac{i\sinh(2\psi )p_\psi }
{ \sqrt{({\cal H}+{\cal L}_2-\gamma )^2+4{\cal L}_2\gamma }}.
\end{gather*}
Thus this system is superintegrable for all rational $k$.

\medskip\noindent {\bf [S7]}. This system corresponds to $v_1(\phi)=Z_2$ and $v_2(\psi)$ a variant of $Z_3$.
The system is second order superintegrable:
\[ V= \frac{\alpha s_3}{ \sqrt {s_1^2+s_2^2}} +\frac {\beta s_1}{ s_2^2\sqrt {s_1^2+s_2^2}} +
\frac{\gamma }{ s_2^2}.\]
Choosing the coordinates $\psi $ and $\varphi $ we f\/ind
\begin{gather*}
 {\cal H}=\cosh^2\psi \left(p^2_\psi +p^2_\varphi + \alpha  \frac{\sinh\psi }{ \cosh^2\psi } + \beta
\frac{\cos\varphi }{ \sin ^2\varphi }+ \frac{\gamma }{ \sin ^2\varphi } \right),\\
 {\cal L}_2=p^2_\varphi  + \beta  \frac{\cos\varphi }{ \sin ^2\varphi }+
\frac{\gamma }{ \sin ^2\varphi },\qquad {\cal H}=\cosh^2\psi \left(p^2_\psi +{\cal L}_2+\alpha \frac {\sinh\psi }{ \cosh^2\psi }\right).
\end{gather*}
We  make the transformation $\varphi \rightarrow k\varphi $ and obtain
\begin{gather*}
{\cal H}=\cosh^2\psi \left(p^2_\psi +p^2_\varphi + \alpha \frac {\sinh\psi}{ \cosh^2\psi } + \beta
\frac{\cos k\varphi }{ \sin ^2k\varphi }+ \frac{\gamma }{ \sin ^2k\varphi } \right ),\\
 {\cal L}_2=p^2_\varphi  + \beta \frac {\cos k\varphi }{ \sin ^2k\varphi }+
\frac{\gamma }{\sin ^2k\varphi },\nonumber
\end{gather*}
with $\cal H$ as before. The functions that determine the extra constants are
\begin{gather*}
 M(\varphi ,p_\varphi )=\frac{1}{ \sqrt {{\cal L}_2}k}{\cal  A},\\ \sinh{\cal  A }=
\frac{i({\cal L}_2\cos (k\varphi )+\beta )}{ \sqrt {\beta ^2+4{\cal L}^2_2-4{\cal L}_2\gamma }},\qquad
\cosh{\cal A}= \frac{2\sqrt{ {\cal L}_2}\sin (k\varphi )p_\varphi }{ \sqrt {\beta ^2+4{\cal L}^2_2-4{\cal L}_2\gamma }},\\
 N(\psi ,p_\psi )=\frac{i}{ \sqrt{{\cal  L}_2}}{\cal B},\\ \sinh{\cal B}= \frac{2{\cal L}_2\sinh\psi +\alpha }{ \sqrt {-\alpha ^2+4{\cal L}^2_2-4{\cal L}_2{\cal H}}},\qquad
\cosh{\cal  B }= \frac{2i\cosh\psi p_\psi }{ \sqrt {-\alpha ^2+4{\cal L}^2_2-4{\cal L}_2{\cal H}}}.
\end{gather*}
Thus this system is superintegrable for all rational~$k$.

\medskip\noindent {\bf [S4]}. Here $v_1(\varphi)=Z_4$ and $v_2(\psi)$ is a variant of $Z_2$.
This is another system on the sphere that is second order superintegrable and  separates in polar coordinates:
\[ V=\frac{\alpha }{ (s_1-is_2)^2} + \frac{\beta s_3}{ \sqrt {s_1^2+s_2^2}} +
\frac{\gamma }{ (s_1-is_2)\sqrt {s_1^2+s_2^2}}.\]
In terms of angular coordinates $\psi $, $\varphi $ the Hamiltonian is
\begin{gather*}
{\cal H}=\cosh^2\psi \left(p^2_\psi +p^2_\varphi +
\alpha e^{2ik\varphi }+\gamma ^{ik\varphi }+\beta \frac {\sinh\psi }{ \cosh^2\psi}\right ).
\end{gather*}
After the substitution $\varphi \rightarrow k\varphi$ we have
\[ {\cal L}_2=p^2_\varphi  + \alpha e^{2ik\varphi }+\gamma ^{ik\varphi },\qquad {\cal H}=\cosh^2\psi \left(p^2_\psi + {\cal L}_2+\beta \frac {\sinh\psi }{ \cosh^2\psi } \right).\]
The functions that determine the extra constants are{\samepage
\begin{gather*}
 M(\varphi ,p_\varphi )=\frac{i}{ 2\sqrt{ {\cal L}_2}k} {\cal A},\qquad
\sinh{\cal A}= \frac{i(2{\cal L}_2e^{-ik\varphi }-\gamma) }{ \sqrt {4{\cal L}_2\alpha +\gamma ^2}},\qquad
\cosh{\cal  A}= \frac{2i\sqrt{ {\cal L}_2}p_\varphi }{ \sqrt {4{\cal L}_2\alpha +\gamma ^2}}
e^{-ik\varphi },\\
 N(\psi ,p_\psi )=\frac{i}{ 2\sqrt{ {\cal L}_2}}{\cal B},\qquad \sinh{\cal B }= \frac{{\cal L}_2\sinh\psi -\beta }{ \sqrt{4{\cal L}^2_2+4{\cal L}_2-\beta ^2}},\qquad
\cosh{\cal  B }= \frac{2i\cosh\psi p_\psi }{ \sqrt {4{\cal L}^2_2+4{\cal L}_2-\beta ^2}},
\end{gather*}
so this systems is also superintegrable for all rational $k$.}

\medskip\noindent {\bf [S2]}.
Here $v_1(\varphi)=Z_4$ and $v_2(\psi)$ is a variant of $Z_3$:
\[ V=\frac{\alpha }{ s_3^2}+ \frac{\beta }{ (s_1-is_2)^2}+ \frac{\gamma (s_1+is_2)}{ (s_1-is_2)^3}.\]
which is is second order superintegrable.
After the substitution $\varphi \rightarrow k\varphi $ we obtain the system
\begin{gather}\label{hamS2} {\cal H}=\cosh^2\psi \left(p^2_\psi +p^2_\varphi + \frac{\alpha }{ \sinh^2\psi }
+\beta e^{2ik\varphi}+\gamma e^{4ik\varphi}\right),\\
 {\cal L}_2=p^2_\varphi +\beta e^{2ik\varphi }+\gamma e^{4ik\varphi },\qquad
{\cal H}=\cosh^2\psi \left(p^2_\psi +{\cal L}_2+ \frac{\alpha }{ \sinh^2\psi }\right).\nonumber
\end{gather}
Applying our procedure we f\/ind
\begin{gather*}
 M(\varphi ,p_\varphi )=\frac{i}{ 4\sqrt{{\cal L}_2}k}{\cal  A},\qquad \sinh{\cal A}= \frac{(2{\cal L}_2e^{-ik\varphi }-\beta )}{ \sqrt {4L_2\gamma -\beta ^2}},\qquad
\cosh{\cal  A }= \frac{2\sqrt{{\cal L}_2}p_\varphi }{ \sqrt {4{\cal L}_2\gamma -\beta ^2}} e^{-ik\varphi },\\
 N(\psi ,p_\psi )=\frac{i}{ 4\sqrt{{\cal  L}_2}}{\cal B},\\ \sinh{\cal  B}=
\frac{i({\cal L}_2\cosh(2\psi )-\alpha +{\cal H})}{ \sqrt {({\cal L}_2-\alpha -{\cal H})^2-4\alpha {\cal H}}},\qquad
\cosh{\cal B }= \frac{\sqrt{{\cal L}_2}\sinh(2\psi )p_\psi }{ \sqrt {({\cal L}_2-\alpha -{\cal H})^2-4\alpha {\cal  H}}}.
\end{gather*}
Thus the system is superintegrable for all rational $k$.

\subsection{Horospherical systems}
In terms of horospherical coordinates on the complex sphere  we can construct the Hamiltonian
\[
 {\cal H}=y^2\left(p^2_x+p^2_y+\omega_1^2x^2+\omega_2^2y^2+\alpha +\beta x\right).
\]
If $\omega_1/\omega_2$ is rational then this system is superintegrable. However, there is no need to go into much detail for  the construction because a St\"ackel transform, essentially multiplication by $1/y^2$, takes this system to the f\/lat space system generalizing [E2] and with the same symmetry algebra.

There is a second Hamiltonian which separates on the complex 2 sphere
\[ H=y^2\left(p^2_x+p^2_y+\frac{\alpha }{ x^2}+\beta +\omega ^2_1x^2+\omega ^2_2y^2\right).
\]
It is superintegrable for  $\omega_1/\omega_2$ rational, as follows from the St\"ackel transform $1/y^2$ from the sphere to f\/lat space.
If $\omega _1=\omega _2$ then
we obtain the system [S2]. We note that  each of the potentials associated with
horospherical coordinates can quite generally be written in terms of~$s_1$, $s_2$, $s_3$
 coordinates using the relations
\[ y=\frac{-i}{ s_1-is_2},\qquad x=\frac{-s_3}{ s_1-is_2},\qquad s_1^2+s_2^2+s_3^2=1.\]

\subsection{Generic  systems}
These are systems of the form
\[
{\cal H} = \frac{1}{ Z_j(x)-Z_k(y)} \big(p^2_x-p^2_y+\hat Z_j(x)-\hat Z_k(y)\big),
\]
where $j$, $k$ can independently take the values  2, 3, 4  and $Z_j$, ${\hat Z}_j$ depend on distinct parameters. Superintegrability is possible  because of the integrals
\begin{gather*}
 \int  \frac{dx}{ \sqrt{C+Be^{2ikx}+Ae^{4ikx}}} = \frac{1}{ 2k\sqrt{ C}} \arcsin\left(\frac{2Ce^{-2ikx}+B}{ \sqrt{4AC-B^2}}\right),\\
 \int \frac{dx}{\sqrt{\frac{A+B\sin kx}{ \cos ^2kx} +C}} = -\frac{1}{ k\sqrt{ C}}
\arcsin\left( \frac{-2C\sin kx+B}{\sqrt{ 4C^2+4CA+B^2}}\right),\\
 \int  \frac{dx}{\sqrt{\frac{A}{ \cos ^2kx} + \frac{B}{ \sin ^2kx} +C}} =
-\frac{1}{ 2k\sqrt{ C}} \arcsin\left( \frac{C\cos 2kx+A-B}{ \sqrt{ (A+B+C)^2-4AB}}\right).
\end{gather*}
Consider an example of this last type of system:
\[ {\cal H} =
\frac{1}{ A'e^{2ikx}+B'e^{4ikx}-a'e^{2iqy}-b'e^{4iqy}}\big(p^2_x-p^2_y+Ae^{2ikx}+Be
^{4ikx}-ae^{2iqy}-be^{4iqy}\big).\]
The equation ${\cal H}=E$ admits a separation constant
\[ p^2_x+(A-EA')e^{2ikx}+(B-EB')e^{4ikx}=p^2_y+(a-Ea')e^{2iqy}+(b-Eb')e^{4iqy}={\cal L}.
\]
As a consequence of this we can f\/ind functions $M(x,p_x)$ and $N(y,p_y)$ where
\[ M(x,p_x)=\frac{1}{ 4k\sqrt{{\cal L}}}\arcsin\left(\frac{2\hat Ae^{-2ikx}+\hat B}{ \sqrt {4L\hat A-\hat B^2}}\right),\]
where $\hat A=A-EA'$, $\hat B=B-EB'$ and
\[ N(y,p_y)=\frac{1}{ 4p\sqrt{ {\cal L}}}
\arcsin \left(\frac{2\hat ae^{-2iqy}+\hat b}{ \sqrt{4{\cal L}\hat a-\hat b^2}}\right),\]
where $\hat a=a-Ea'$, $\hat b=b-Eb'$. We see that if $\frac{q}{ k}$ is rational
then we can generate an extra constant which is polynomial in the momenta.

The superintegrable systems that have involved the rational functions $Z_1$, $Z_5$
and $Z_6$ work because of the integrals
\[ \int  \frac{dx}{ \sqrt{Ax^2+Bx+C}} = \frac{1}{\sqrt { A}} \, {\rm arcsinh} \left(\frac{2Ax+C}{\sqrt{ 4AB-C^2}} \right)
\]
and
\[
\int  \frac{dx}{ x\sqrt {Ax^2+Bx+C}} = -\frac{1}{ \sqrt{ B}} \, {\rm arcsinh}\left(\frac{2B+Ax}{ x\sqrt {A^2-4BC}} \right).
\]

\section{Quantum superintegrability}\label{section2}

\subsection{The canonical form for a symmetry operator}

We give a brief review of the construction of the canonical form for a symmetry operator  \cite{KKM10a}.
Consider a Schr\"odinger equation on a 2D real or complex Riemannian manifold with Laplace--Beltrami operator $\Delta_2$ and potential $V$:
\begin{gather*}
H\Psi\equiv (\Delta_2+V)\Psi=E\Psi
\end{gather*}
that also  admits an orthogonal  separation of variables.
If $\{u_1,u_2\}$ is the  orthogonal separable coordinate system   the corresponding Schr\"odinger
operator has the form
\begin{gather}
H= L_1= \Delta_2+V(u_1,u_2)=
\frac{1}{f_1(u_1)+f_2(u_2)}\left(\partial^2_{u_1}+\partial^2_{u_2}+v_1(u_1)+v_2(u_2)
\right).\label{TIS1}
\end{gather}
and, due to the separability, there is the second-order symmetry operator
\[
L_2= \frac{f_2(u_2)}{f_1(u_1)+f_2(u_2)}\left(\partial^2_{u_1}+v_1(u_1)\right)
-\frac{f_1(u_1)}{f_1(u_1)+f_2(u_2)}\left(\partial^2_{u_2}+v_2(u_2)\right),
\]
i.e., $
[L_2,H]=0,$
and the operator identities
\begin{gather*}
f_1(u_1)H+L_2=\partial^2_{u_1}+v_1(u_1),\qquad
f_2(u_2)H-L_2=\partial^2_{u_2}+v_2(u_2).
\end{gather*}

We  look for a  partial dif\/ferential  operator  ${\tilde L}(H,L_2,u_1,u_2)$ that satisf\/ies
\begin{gather*}
[H,{\tilde L}]= 0.
\end{gather*}
We require that the symmetry operator take the standard form
\begin{gather}\label{standardLform}
{\tilde
L}=\sum_{j,k}\big(A^{j,k}(u_1,u_2)\partial_{u_1u_2}+B^{j,k}(u_1,u_2)\partial_{u_1}
+C^{j,k}(u_1,u_2)\partial_{u_2}+ D^{j,k}(u_1,u_2)
\big)H^jL_2^k.
\end{gather}

We have shown that  we can write
\begin{gather}
{\tilde L}(H,L_2,u_1,u_2)=A(u_1,u_2,H,L_2)\partial_{12}+B(u_1,u_2,H,L_2)\partial_{1}\nonumber\\
\phantom{{\tilde
L}(H,L_2,u_1,u_2)=}{} +C(u_1,u_2,H,L_2)\partial_{2}
+ D(u_1,u_2,H,L_2),\label{generalLform}
\end{gather}
and consider $\tilde  L$ as an at most second-order  order dif\/ferential operator in $u_1$, $u_2$ that is analytic in the parameters $H$, $L_2$.
Then the conditions for a symmetry  can be written in the compact form
\begin{gather} \label{partialxyHL}
A_{u_1u_1}+A_{u_2u_2}+2B_{u_2}+2C_{u_1}
=0,
\\
 \label{partialxHL}
B_{u_1u_1}+B_{u_2u_2}-2A_{u_2}v_2+2D_{u_1}-Av'_2+(2A_{u_2}f_2+Af'_2)H-2A_{u_2}L_2=0,
\\
 \label{partialyHL}
C_{u_1u_1}+C_{u_2u_2}-2A_{u_1}v_1+2D_{u_2}-Av'_1+(2A_{u_1}f_1+Af'_1)H+2A_{u_1}L_2=0,
\\
D_{u_1u_1}+D_{u_2u_2}-2B_{u_1}v_1-2C_{u_2}v_2-Bv'_1-Cv'_2\nonumber
\\
\qquad {} +(2B_{u_1}f_1+2C_{u_2}f_2+Bf'_1
+Cf'_2)H+(2B_{u_1}-2C_{u_2})L_2=0.\label{constanttermHL}
\end{gather}

We can further simplify this system by noting  that there are two functions $F(u_1,u_2,H,L_2)$, $G(u_1,u_2,H,L_2)$ such
that (\ref{partialxyHL}) is satisf\/ied by
\begin{gather}\label{ABC}
A=F,\qquad B=-\frac12 \partial_{2}F-\partial_{1}G,\qquad C=-\frac12\partial_{1} F+\partial_{2}G.
\end{gather}
Then the integrability condition for (\ref{partialxHL}), (\ref{partialyHL}) is (with the shorthand $\partial_{j}F=F_j$, $\partial_{j\ell}F=F_{j\ell}$, etc., for $F$ and $G$),
\begin{gather}
2G_{1222}+\frac12 F_{2222}+2F_{22}(v_2-f_2H+L_2)
+3F_{2}(v'_2-f_2'H)+F(v''_2-f''_2H) \nonumber\\
\qquad{} -2G_{1112} +\frac12 F_{1111}+2F_{11}(v_1-f_1H-L_2)
+3F_{1}(v'_1-f'_1H)+F(v''_1-f''_1H),\label{eqn1}
\end{gather}
and equation (\ref{constanttermHL}) becomes
\begin{gather} \frac12
F_{1112}+2F_{12}(v_1-f_1H)+F_{1}(v'_2-f'_2H)+\frac12
G_{1111}+ 2G_{11}(v_1-f_1H-L_2)\nonumber\\
\qquad{} +G_{1}(v'_1-
f'_1H)
=-\frac12F_{1222}-2F_{12}(v_2-f_2H)-F_{2}(v'_1-f'_1H)+\frac12
G_{2222} \nonumber\\
\qquad{}+ 2G_{22}(v_2-f_2H+L_2)+G_{2}(v'_2-f'_2H). \label{eqn2}
\end{gather}
Here,  any solution of (\ref{eqn1}), (\ref{eqn2}) with~$A$, $B$, $C$ not identically $0$ corresponds to a symmetry operator that does not commute with~$L_2$, hence is algebraically  independent of the symmet\-ries~$H$,~$L_2$. (Informally, this follows from the construction and uniqueness of the canonical form of a symmetry operator. The operators ${\tilde L}=g(H,L_2)$ algebraically dependent on $H$ and $L_2$ are exactly those such that $A=B=C=0$, $D=g(H,L_2)$. A formal proof is technical.)  Note also that solutions of the canonical equations, with $H$, $L_2$ treated as parameters, must be interpreted in the form (\ref{standardLform}) with $H$ and $L_2$ on the right, to get the explicit symmetry operators.

\subsection{The caged anisotropic oscillator}\label{cagedosc}

In \cite{KKM10a} we used the canonical form for symmetry operators to demonstrate the quantum superintegrability of the TTW system \cite{TTW, TTW2} for all rational $k$, as well as a system on the 2-hyperboloid of two sheets \cite{KKM10a}. Here we give another illustration of this construction by applying it to
the 2D caged oscillator  \cite{Evans2008a}. For $p=q$ This is the second order superintegrable system [E1] on complex Euclidean space, as listed in \cite{KKMP}.  Here,
\begin{gather}\label{cagedoscillator}
H\Psi=(\partial_{11}+\partial_{22} +V(u_1,u_2))\Psi,\end{gather}
where
\[ V(u_1,u_2)=\omega^2\big(p^2u_1^2+q^2u_2^2\big)+\frac{\alpha_1}{u_1^2}+\frac{\alpha_2}{u_2^2},\]
in Cartesian coordinates. We take $p$, $q$ to be relatively prime positive integers.
Thus
\begin{gather*}
f_1=1,\qquad f_2=0,\qquad v_1=\omega^2 p^2u_1^2+\frac{\alpha_1}{u_1^2},\qquad v_2=\omega^2 q^2u_2^2+\frac{\alpha_2}{u_2^2}.
\end{gather*}
The 2nd order symmetry operator is
\[
L_2= -\left(\partial^2_2+v_2(u_2)\right),
\]

and we have the operator identities
\begin{gather*} 
\partial^2_1=-(v_1(u_1)+H+L_2),\qquad
\partial^2_2=-v_2(u_2)+L_2.
\end{gather*}

Based on the results of \cite{KKM10} for the classical case, we postulate expansions of  $F$, $G$ in f\/inite series
\begin{gather}\label{TTW1} F=\sum_{a,b} A_{a,b}E_{a,b}(u_1,u_2),\qquad G=\sum_{a,b} B_{a,b}E_{a,b}(u_1,u_2),\qquad E_{a,b}(u_1,u_2)=u_1^au_2^b.
\end{gather}

The sum is taken over terms of the form $a=a_0+m$, $b=b_0+n$, and $c=0,1$, where~$m$,~$n$ are integers.  The point $(a_0,b_0)$ could in principle be any point in $\mathbb R^2$,

Taking coef\/f\/icients with respect to the basis (\ref{TTW1}) in each of equation (\ref{eqn1}) and (\ref{eqn2})
gives recurrence relations for these coef\/f\/icients.
The shifts in the indices of $A$ and $B$ are integers and so we can view this as an equation on a two-dimensional
lattice with integer spacings.  While the shifts in the indices are of integer size, we haven't required that the
indices themselves be integers, although they will turn out to be so in our solution.
The 2 recurrence relations are of a similar complexity, but rather than write them
out separately, we will combine them into a~matrix recurrence relation by
def\/ining
\[
 {C}_{a,b} = \left( \begin{array}{c} A_{a,b} \\ B_{a,b}  \end{array} \right).
\]
We  write the 2 recurrence relations in matrix form as
\begin{gather*}
 \mathbf{0} = {M}_{a,b}{C}_{a,b}  +
 {M}_{a-1,b+1}{C}_{a-1,b+1} +
 {M}_{a,b+2}C_{a,b+2} +
 {M}_{a,b+4}C_{a,b+4} \nonumber \\
\phantom{\mathbf{0} =}{} +
 {M}_{a+1,b-1}{C}_{a+1,b-1} +
 {M}_{a+1,b+1}{C}_{a+1,b+1} +
 {M}_{a+1,b+3}C_{a+1,b+3} +
 {M}_{a+2,b}{C}_{a+2,b} \nonumber \\
\phantom{\mathbf{0} =}{}   + {M}_{a+3,b+1}{C}_{a+3,b+1}+{M}_{a+4,b}{C}_{a+4,b}, 
\end{gather*}
where each ${M}_{i,j}$ is a $2\times2$ matrix given below.
\begin{gather*}
{M}_{a,b} =
 \left( \begin {array}{cc}
 2\omega^2\left([b+1]q-[a+1]p\right)\left([b+1]q+[a+1]p\right)
 & 0 \\
 0
 & -2\omega^2\left(bq-ap\right)\left(bq+ap\right)
 \end {array} \right),
\\
{M}_{a-1,b+1} =
 \left( \begin {array}{cc}
 0
 & 0 \\
 2\omega^2p^2a(b+1)
 & 0
 \end {array} \right),
\\
{M}_{a,b+2} =
 \left( \begin {array}{cc}
 2L_2(b+2)(b+1)
 & 0 \\
 0
 & -2L_2(b+2)(b+1)
 \end {array} \right),
\\
{M}_{a,b+4} =
 \left( \begin {array}{cc}
 \frac12(b+3)(b+1)(b^2+6b+4\alpha_2+8)
 & 0 \\
 0
 & -\frac12(b+4)(b+2)(b^2+4b+4\alpha_2+3)
 \end {array} \right),
\\
{M}_{a+1,b-1} =
 \left( \begin {array}{cc}
 0
 & 0 \\
 2\omega^2q^2b(a+1)
 & 0
 \end {array} \right),
\\
{M}_{a+1,b+1} =
 \left( \begin {array}{cc}
 0
 & 0 \\
 -2H(a+1)(b+1)
 & 0
 \end {array} \right),
\\
{M}_{a+1,b+3} =
 \left( \begin {array}{cc}
 0
 & 2(a+1)(b+3)(b+2)(b+1)\\
\frac12 (a+1)(b+2)(b^2+4b+4\alpha_2+3)
 & 0
 \end {array} \right),
\\
{M}_{a+2,b} =
 \left( \begin {array}{cc}
 2(H+L_2)(a+2)(a+1)
 & 0 \\
 0
 & -2(H+L_2)(a+2)(a+1)
 \end {array} \right),
\\
{M}_{a+3,b+1} =
 \left( \begin {array}{cc}
 0
 & 2(a+3)(a+2)(a+1)(b+1)\\
\frac12 (a+2)(b+1)(a^2+4a+4\alpha_1+3)
 & 0
 \end {array} \right),
\\
{M}_{a+4,b} =
 \left( \begin {array}{cc}
 -\frac12(a+3)(a+1)(a^2+6a+4\alpha_1+8)
 & 0 \\
 0
 & \frac12(a+4)(a+2)(a^2+4a+4\alpha_1+3)
 \end {array} \right).\!
\end{gather*}

\begin{figure}[t]
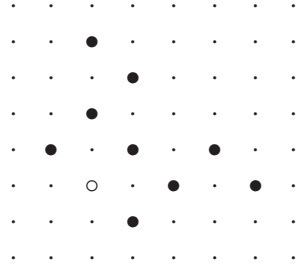

$$
 \begin{array}{ccccccccccc}
   \cdot & \cdot   & \cdot   & \cdot   &\cdot  &\cdot   & \cdot & \cdot \\
   \cdot & \cdot   & \bullet & \cdot   &\cdot  &\cdot   & \cdot & \cdot \\
   \cdot & \cdot   & \cdot   &\bullet  &\cdot  &\cdot & \cdot & \cdot  \\
   \cdot & \cdot   & \bullet &\cdot    &\cdot  &\cdot   & \cdot & \cdot  \\
   \cdot & \bullet & \cdot   &\bullet  &\cdot  &\bullet  & \cdot & \cdot  \\
   \cdot & \cdot   & \circ &\cdot    &\bullet&\cdot   &\bullet & \cdot \\
   \cdot & \cdot   & \cdot   &\bullet  &\cdot  &\cdot   & \cdot & \cdot  \\
   \cdot & \cdot   & \cdot   &\cdot    &\cdot  &\cdot   & \cdot & \cdot
 \end{array}
$$
\caption{The template. Points contributing to the recurrence relation are marked with large dots~($\bullet$,~$\circ$).  The large dot
on the bottom center
 corresponds to the position $(a-1,b+1)$, the large dot at the top corresponds to the position $(a+4,b)$ and $\circ$ corresponds to position $(0,0)$.}
\label{template}
\end{figure}

It is useful to visualize the
the set of points in the lattice which enter into this recurrence for a given choice of $(a,b)$.
These are represented in Fig.~\ref{template}. Although the recurrence relates $10$~distinct points, some major simplif\/ications are immediately apparent. We say that the lattice point  $(a,b)$ has even parity if $a+b$ is an even integer and odd parity if $a+b$ is odd. Each recurrence relates only lattice points of the same parity. Because of this we can assume that the nonzero  terms $C_{a,b}$ will occur for points of one parity, while only the zero vector will occur for  points with the opposite parity.  A second simplif\/ication results from the fact that the recurrence matrices $M_{a+m,b+n}$ are of two distinct types. Either $m$, $n$ are both even, in which case $M_{a+m,b+n}$ is diagonal, or $m$, $n$ are both odd, in which case the diagonal elements of $M_{a+m,b+n}$ are zero. Another simplif\/ication follows from the observation that it is only the ratio $p/q=r$ that matters in our construction. We want to demonstrate that the caged anisotropic oscillator is operator superintegrable for any rational~$r$. The construction of any  symmetry operator independent of~$H$ and~$L_2$ will suf\/f\/ice. By writing $\omega=\omega'/2$, $p'=2p$, $q'=2q$ in~$H$, we see that, without loss of generality, we can always assume that $p$, $q$ are both even positive integers with a single~2 as their only common factor.

If for a particular choice of $p$, $q$ we can f\/ind a solution of the recurrence relations with only a~f\/inite number of nonzero vectors $C_{a,b}$ then there will be a minimal lattice rectangle with vertical sides and horizontal top and bottom that encloses the corresponding lattice points. The top row~$a_0$ of the rectangle will be the highest row in which nonzero vectors $C_{a_0,b}$ occur. The bottom row~$a_1$ will be the lowest row in which nonzero vectors $C_{a_1,b}$ occur. Similarly, the minimal rectangle will have right column $b_0$ and left column $b_1$.  Now slide the template horizontally across the top row such that only the lowest point on the template lies in the top row. The recurrence gives $(a_0+1)bA_{a_0,b}=0$ for all columns $b$. Based on examples for the classical system, we expect to f\/ind solutions for the quantum system such that $a_0\ge 0$ and $b\ge 0$. Thus we require $A_{a_0,b}=0$ along the top of the minimal lattice rectangle, so that all vectors in the top row take the form
\[ C_{a_0,b}=\left(\begin{array}{c} 0\\ B_{a_0,b}\end{array}\right).\]
 Now we move the template such that the recurrence $M_{a,b}$, second row from the  bottom and on the left, lies on top of the lattice point $(a_0,b_0)$. This leads to the requirement $(b_0q-a_0p)(b_0q+a_0p)B_{a_0,b_0}=0$. Again, based on hints from specif\/ic examples, we postulate $B_{a_0,b_0}\ne 0$, \mbox{$a_0=q$}, $b_0=p$. Since we can assume  that $p$, $q$ are even, this means that all odd lattice points correspond to zero vectors. Next we   slide the template vertically down  column~$b_0$ such that only the left hand  point on the template lies in column~$b_0$. The recurrence gives $(b_0-1)aA_{a,b_0}=0$ for all rows~$a$. Since $b_0$ is even, we postulate that
\[  C_{a,b_0}=\left(\begin{array}{c} 0\\ B_{a,b_0}\end{array}\right)\]
for all lattice points in column $b_0$.

Note that the recurrence relations preserve the following structure which we will require:
\begin{enumerate} \itemsep=0pt
\item There is only the zero vector at any lattice point $(a,b)$ with $a+b$  odd. \item If the row and column are both even then
$C_{a,b}=\left(\begin{array}{c} 0\\B_{a,b}\end{array}\right)$. \item  If the row and column are both odd then
$C_{a,b}=\left(\begin{array}{c} A_{a,b}\\ 0\end{array}\right)$.
\end{enumerate}
 This does not mean that all solutions have this form, only that we are searching for  at least one such solution.

To f\/inish determining the size of the minimum lattice rectangle we   slide the template vertically downward such that only the right hand  point on the template lies in column $b_1$. The recurrence gives $(b_1-1)(b_1-3)A(a,b_1)=0$, $b_1(b_1-2)B_{a,b_1}=0$ for all rows $a$. There are several possible solutions that  are in accordance with our assumptions, the most conservative of which is $b_1=0$. In the following we assume only that $B_{a_0,b_0}\ne 0$, the structure laid out above, and the necessary implications of these that follow from a step by step application of the recurrences. If we f\/ind a~vector at a lattice point that is not determined by the recurrences we shall assume it to be zero. Our aim is to f\/ind one nonzero solution with support in the minimum rectangle, not to classify the multiplicity of all such solutions.

Now we carry out  an iterative procedure that calculates the values of
${C}_{a,b}$ at points in the lattice using only other points where
the values of ${C}_{i,j}$ are already known.   We position the template such that the recurrence $M_{a,b}$, second row from the  bottom and on the left (of the template), lies above  the lattice point $(a_0,b)$, $a_0=q$, and slide it from right to left along the top  row. The case $b=b_0=p$ has already been considered. For the remaining cases we have
\begin{gather} 2\omega^2p^2a_0(b+1) A_{a_0-1,b+1} =2\omega^2(bq-a_0p)(bq+a_0p)B_{a_0,b}+2L_2(b+2)(b+1)B_{a_0,b+2}\nonumber\\
\qquad{} +\frac12 (b+4)(b+2)(b^2+4b+4\alpha_2+3)B_{a_0,b+4}.\label{1strow}
\end{gather}
Only even values of $b$ need be considered.  Now we lower the template one row and again slide it from right to left along the row. The contribution of the lowest point on the template is $0$ and we f\/ind
\begin{gather}  2\omega^2(q[b+2]-p[a_0]) (q[b+2]+p[a_0])A_{a_0-1,b+1} =-2a_0(b+4)(b+3)(b+2)B_{a_0,b+4} \nonumber\\
\qquad{} -2L_2(b+3)(b+2)A_{a_0-1,b+3}-\frac12 (b+4)(b+2)(b^2+8b+11\alpha_2+8)A_{a_0-1,b+5}, \label{2ndrow}
\end{gather}
where we have replaced $b$ by $b+1$ so that again only even values of $b$ need be considered.

For the f\/irst step we take $b=p-2$. Then equation (\ref{1strow}) becomes
\[  2\omega^2p^3qA_{q-1,p-1}=-8\omega^2q^2(p-1)B_{q,p-2}+2L_2p(p-1)B_{q,p},\]
 and (\ref{2ndrow}) is vacuous in this case, leaving  $A_{q-1,p-1}$ undetermined.
Note that   several of the terms lie outside the minimal rectangle. From these two equations we can solve for $B_{q,p-2}$ in terms of our given $B_{q,p}$, $A_{q-1,p}$. Now we march across both rows from right to left. At each stage our two equations now allow us to solve uniquely for $A_{q-1,b+1}$ and $B_{q,b}$ in terms of $A$'s and $B$'s to the right (which have already been computed). We continue this until we reach $b=0$ and then stop. At this point the top two rows of the minimal rectangle have been determined by our choice of~$B_{q,p}$ and $A_{q-1,p-1}$. We repeat this construction for the third and fourth  rows down, then the next two rows down, etc. The recurrence relations grow more complicated as the higher rows of the template give nonzero contributions. However, at each step we have two linear relations
\[  \left(\begin{array}{cc} 2\omega^2p^2a(b+1) &-2\omega^2(bq-ap)(bq+ap)\\ 2\omega^2(q[b+2]-pa)(q[b+2]+pa) & 0\end{array}\right)\left(\begin{array}{c} A_{a-1,b+1}\\ B_{a,b}\end{array}\right)=\cdots, \]
 where the right hand side is expressed in terms of $A$'s and $B$'s either above or on the same line but to the right of $A_{a-1,b+1}$, $B_{a,b}$, hence already determined. Since the determinant of the $2\times 2$ matrix is nonzero over the minimal rectangle, except for the upper right corner and lower left corner,  at all but those those two points  we can  compute $A_{a-1,b+1}$, $B_{a,b}$ uniquely in terms of quantities already determined. This process stops with rows $a=1,2$.

Row $a=0$, the bottom row, needs special attention. We position the template such that the recurrence $M_{a,b}$  lies above  the lattice point $(0,b)$,  and slide it from right to left along the bottom row. The bottom point in the template now contributes $0$ so at each step we obtain an expression for $B_{0,b}$ in terms of quantities $A$, $B$ from rows either above row $0$ or to the right of column $b$ in row $0$, hence already determined. Thus we can determine the entire bottom row, with the exception of the value at $(0,0)$, the lower left hand corner. When the point $M_{a,b}$ on the template is above $(0,0)$ the coef\/f\/icient of $B_{0,0}$ vanishes, so the value of $B_{0,0}$ is irrelevant and we have a true condition on the remaining points under the template. However, this is a linear homogeneous equation in the parameters $B_{q,p}$, $A_{q-1,p-1}$: $\chi B_{q,p} +\eta A_{q-1,p-1}=0$ for constants~$\chi$,~$\eta$. Hence  if, for example,  $\chi\ne 0$ we can require $B_{q,p} =-(\eta/\chi) A_{q-1,p-1}$ and satisfy this condition while still keeping a nonzero solution. Thus after satisfying this linear condition we still have at least a one parameter family of solutions, along with the arbitrary $B_{0,0}$ (which is clearly irrelevant since it just adds a constant to the function $G$). However, we need to check those recurrences where the point $M_{a,b}$ on the template slides along rows $a=-1,-2,-3,-4$ to verify that these relations are satisf\/ied. We have already utilized the case $a=-4$, and for cases  $a=-3,-2,-1$ it is easy to check that the relations are vacuous.

The last issue is the left hand boundary. We have determined all vectors in the minimal rectangle, but we must verify that those recurrences are satisf\/ied where the point $M_{a,b}$ on the template slides along columns $b=-1,-2,-3,-4$.  However again the recurrences are vacuous.

We conclude that there is a one-parameter (at least) family of solutions to the caged ani\-so\-tro\-pic oscillator recurrence with support in the minimal rectangle. By choosing the arbitrary parameters to be polynomials in~$H$,~$L_2$ we get a f\/inite order constant of the motion. Thus the quantum caged anisotropic oscillator is superintegrable.  We note that the canonical operator construction permits easy generation of explicit expressions for the def\/ining operators in a large number of examples. Once the basic rectangle of nonzero solutions is determined it is easy to compute  dozens of explicit examples via Maple and simple Gaussian elimination. The generation of explicit examples is easy; the proof that the method works for all orders is more challenging.

\medskip\noindent {\bf Example.}
Taking $p=6$ and $q=4$, we f\/ind (via Gaussian elimination) a solution of the recurrence with nonzero coef\/f\/icients
\begin{gather*}
 A_{1,1}, \ A_{1,3}, \ A_{1,5}, \ A_{3,1}, \ A_{3,3}, \ A_{3,5},
\\
 B_{0,2},\ B_{0,4},\ B_{2,0},\ B_{2,2}, \ B_{2,4}, \ B_{4,0}, \ B_{4,2}, \ B_{4,4}, \ B_{6,0}, \ B_{6,2},\ B_{6,4}.
\end{gather*}
We can take $A_{1,5}=a_1$ and $A_{3,3}=a_2$ and then all other coef\/f\/icients will depend
linearly on~$a_1$ and~$a_2$.  Calculating the $A$, $B$, $C$ and $D$ from equations (\ref{partialxHL}), (\ref{partialyHL}), (\ref{ABC})
we f\/ind the coef\/f\/icient of~$H^{-1}$ has a factor of $2a_2+9a_1$ and so we set
$a_1=\omega^6$ and $a_2=-9/2a_1$ to obtain the symmetry operator $\tilde L$, (\ref{generalLform}), where $u_1$, $u_2$ are Cartesian coordinates and
\begin{gather*}
A =
 {\frac {9}{4096}}{\omega}^{2}u_1u_2L_2^{2}
+{\frac {3}{2048}}{\omega}^{2}u_1u_2HL_2
-{\frac {1}{256}}{\omega}^{4}u_1u_2 \left(27u_1^{2} -32u_2^{2} \right) L_2 \\
\phantom{A=}{}
+\frac{1}{16}{\omega}^{4}u_1u_2^{3}{H}
+{\frac {1}{64}}{\omega}^{4}{\alpha_2}u_1u_2-\frac{1}{2} {\omega}^{6}u_1u_2^{3} \left( 9u_1^{2}-2u_2^{2} \right)-{\frac {93}{256}}{\omega}^{4}u_1u_2,
\\
B =
-{\frac {1}{786432}}u_1L_2^{4}
-{\frac {1}{786432}}u_1{H}L_2^{3}
+{\frac {3}{32768}}{\omega}^{2}u_1 \left( u_1^2-4u_2^2 \right)  L_2^{3}
-{\frac {3}{8192}}{\omega}^{2}u_1u_2^{2}{H}L_2^{2} \\
\phantom{B=}{}
+ \left( {\frac {149}{49152}}{\omega}^{2}u_1-{\frac {1}{4096}}{\omega}^{2}{\alpha_2}u_1+{\frac {1}{1024}} {\omega}^{4}u_1u_2^{2} \left(27u_1^{2} -16u_2^{2} \right) \right) L_2^{2} \\
\phantom{B=}{}
+ \left( {\frac {59}{49152}}{\omega}^{2}u_1-{\frac {1}{4096}}{\omega}^{2}{\alpha_2}u_1-{\frac {1}{64}}{\omega}^{4}u_1u_2^{4} \right) {H}L_2 \\
\phantom{B=}{}
+ \left( {\frac {45}{512}}{\omega}^{4}u_1u_2^{2}-{\frac {1}{128}}{\omega}^{4}{\alpha_2}u_1u_2^{2}-\frac{1}{6}{\omega}^{6}u_1u_2^{6} \right) {H}+ \left( {\frac {1}{512}} {\omega}^{4}{\alpha_2}u_1 \left( 9u_1^2-4u_2^2 \right)\right.\\
\left.\phantom{B=}{}
 +\frac{1}{24}{\omega}^{6}u_1u_2^{4} \left(27u_1^{2}  -4u_2^{2}\right) -{\frac {3}{2048}} {\omega}^{4}u_1 \left( 59u_1^{2}-284u_2^{2} \right) \right) L_2 \\
\phantom{B=}{}
+{\frac {9}{16}}{\omega}^{6}{\alpha_2}u_2^{2}u_1^{3}+{\frac {17}{128}}{\omega}^{4}{\alpha_2}u_1-{\frac {1}{64}}u_1u_2^{2} \left(405u_1^{2} -416u_2^{2} \right) {\omega}^{6}+12{\omega}^{8}u_1^{3}u_2^{6}-{\frac {45}{512}}{\omega}^{4}u_1,
\\
C =
-{\frac {1}{786432}}u_2L_2^{4}
-{\frac {1}{786432}}u_2{{H}}^{2}L_2^{2}
-{\frac {1}{393216}}u_2{H}L_2^{3}
+{\frac {1}{73728}}{\omega}^{2}u_2 \left( 27u_1^{2}-8u_2^{2} \right) L_2^{3} \\
\phantom{C=}{}
+{\frac {1}{73728}}{\omega}^{2}u_2 \left( 27u_1^{2}-16u_2^{2} \right) {H}L_2^{2}
-{\frac {1}{9216}}{\omega}^{2}u_2^{3}{{H}}^{2}L_2 + \left( {\frac {3}{16384}}{\omega}^{2}{\alpha_1}u_2\right.\\
\left.\phantom{C=}{}
-{\frac {1}{36864}}{\omega}^{2}{\alpha_2}u_2+{\frac {601}{196608}}{\omega}^{2}u_2-{\frac {1}{18432}}u_2 \left( 243u_1^{4}-576u_1^{2}u_2^{2}+32u_2^{4} \right) {\omega}^{4} \right) L_2^{2} \\
\phantom{C=}{}
+ \left( {\frac {85}{24576}}{\omega}^{2}u_2-{\frac {1}{18432}}{\omega}^{2}{\alpha_2}u_2+{\frac {1}{288}} {\omega}^{4}u_2^{3} \left( 9u_1^2-u_2^2 \right) \right) {H}L_2 \\
\phantom{C=}{}
+ \left( {\frac {31}{49152}}{\omega}^{2}u_2-{\frac {1}{36864}}{\omega}^{2}{\alpha_2}u_2-{\frac {1}{576}}{\omega}^{4}u_2^{5} \right) {{H}}^{2} \\
\phantom{C=}{}
+ \left( {\frac {1}{128}}u_1^{2}u_2{\omega}^{4}{\alpha_2}+\frac{1}{2}{\omega}^{6}u_2^{5}u_1^{2}-{\frac {3}{512}}u_2 \left( -16u_2^{2}+31u_1^{2} \right) {\omega}^{4} \right) {H}+ \left( {\frac {1}{128}}{\omega}^{4}{\alpha_2}u_1^{2}u_2\right. \\
\left.\phantom{C=}{}
-{\frac {1}{64}}{\omega}^{4}{\alpha_1}u_2^{3}-\frac{1}{8}u_1^{2}u_2^{3} \left( 9u_1^2-2u_2^2 \right) {\omega}^{6}-{\frac {1}{256}}u_2 \left(114u_1^{2} -29u_2^{2} \right) {\omega}^{4} \right) L_2 \\
\phantom{C=}{}
-{\frac {1}{256}}{\omega}^{4}{\alpha_1}{\alpha_2}u_2-\frac{1}{4}{\omega}^{6}{\alpha_1}u_2^{5}+{\frac {93}{1024}}{\omega}^{4}{\alpha_1}u_2-{\frac {9}{32}}{\omega}^{6}{\alpha_2}u_2u_1^{4}+{\frac {5}{1024}}{\omega}^{4}{\alpha_2}u_2 \\
\phantom{C=}{}
+{\frac {1}{128}} {\omega}^{6}u_2 \left( 40u_2^{4}+837u_1^{4}-1440u_2^{2}u_1^{2} \right)-18{\omega}^{8}u_2^{5}u_1^{4}-{\frac {465}{4096}}{\omega}^{4}u_2,
\\
D =
  {\omega}^{2}\left({\frac {75}{65536}}u_1^{2} -{\frac {55}{49152}}u_2^{2} \right)L_2^{3}
+ {\omega}^{2}\left({\frac {15}{16384}}u_1^{2} -{\frac {83}{49152}}u_2^{2} \right) {H}L_2^{2}
-{\frac {7}{12288}}{\omega}^{2}u_2^{2}{{H}}^{2}L_2 \\
\phantom{D=}{}
+ {\omega}^{4}\left({\frac {471}{2048}}u_2^{2}u_1^{2}-{\frac {5}{144}}u_2^{4}-{\frac {135}{4096}}u_1^{4} \right) L_2^{2}
+{\frac {1}{1152}} {\omega}^{4}u_2^{2} \left( 189u_1^{2}-53u_2^{2} \right){H}L_2
 \\
\phantom{D=}{}
-{\frac {13}{1152}}{\omega}^{4}u_2^{4}{{H}}^{2}
+ \left(   {\omega}^{4}{\alpha_2}\left( {\frac {113}{1024}}u_1^{2}-{\frac {3}{256}}u_2^{2} \right)-{\frac {21}{256}}{\omega}^{4}{\alpha_1}u_2^{2}\right. \\
\left.
\phantom{D=}{}
-\frac{1}{32} {\omega}^{6}u_2^{2} \left( 8u_2^{4}-194u_2^{2}u_1^{2}+189u_1^{4} \right)
+ \left({\frac {15}{64}}u_2^{2} -{\frac {1065}{4096}}u_1^{2} \right) {\omega}^{4} \right) L_2 \\
\phantom{D=}{}
+ \left( {\omega}^{4}{\alpha_2} \left( {\frac {17}{256}}u_1^{2}-{\frac {3}{256}}u_2^{2} \right) +\frac{1}{4} {\omega}^{6}u_2^{4} \left( 13u_1^{2}-u_2^{2} \right)+  {\omega}^{4}\left({\frac {135}{1024}}u_2^{2} -{\frac {45}{1024}}u_1^{2} \right) \right) {H} \\
\phantom{D=}{}
-{\omega}^{6}{\alpha_1}{\frac {13}{8}}u_2^{4}-{\frac {9}{64}} {\omega}^{6}{\alpha_2}u_1^{2} \left( 17u_1^{2}-10u_2^{2} \right)
+  {\omega}^{6}\left( {\frac {405}{256}}u_1^{4}-{\frac {2025}{128}}u_1^{2}u_2^{2}+{\frac {65}{32}}u_2^{4} \right) \\
\phantom{D=}{}
-3u_1^{2} {\omega}^{8}u_2^{4} \left( 39u_1^{2}-10u_2^{2} \right).
\end{gather*}

 Using Maple, we have checked  explicitly that the operator $\tilde L$ commutes with $H$.
 Note that it is of 6th order. Taking the formal adjoint ${\tilde L}^*$, \cite{KKM10a}, we see that $S=\frac12({\tilde L}+{\tilde L}^*)$ is a 6th order, formally self-adjoint symmetry operator.

\section{An alternate proof of  superintegrability\\ for the caged quantum oscillator}\label{section3}

This second proof is very special for the oscillator and exploits the fact that separation of variables in Cartesian coordinates is allowed  Here we write the Hamiltonian in the form
\begin{gather*}
H=\partial ^2_x+\partial ^2_y-\mu_1 ^2x^2-\mu_2 ^2y^2+
\frac{\frac14-a_1^2}{x^2} + \frac{\frac14-a_2^2}{ y^2}.
\end{gather*}
This is the same as (\ref{cagedoscillator}) with $u_1=x$, $u_2=y$, $\mu_1^2=-p^2\omega^2$, $\mu_2^2=-q^2\omega^2$, $\alpha_1=\frac14-a_1^2$, $\alpha_2=\frac14 -a_2^2$.
We look for eigenfunctions for the equation $H\Psi=\lambda\Psi$  of the form $\Psi =XY$.  We f\/ind the normalized solutions
\[ X_n=e^{-\frac12\mu _1x^2}x^{a_1+\frac12} L^{a_1}_n(\mu_1x^2),\qquad
Y_m=e^{-\frac12\mu_2y^2}y^{a_2+\frac12} L^{a_2}_m(\mu _2y^2),
\]
where the $L_n^\alpha(x)$ are associated Laguerre polynomials~\cite{AAR}.
For the corresponding separation constants we obtain
\[ \lambda _x=-2\mu _1(2n+a_1+1),\qquad
\lambda _y=-2\mu _2(2m+a_2+1).\]
The total energy is $-\lambda _x-\lambda _y=E$. Taking
$\mu _1=p\mu $ and $\mu _2=q\mu $ where $p$ and $q$ are integers
we f\/ind that the total energy is
\[
E=-2\mu (pn+qm+pa_1+p+qa_2+q).
\]

Therefore, in order that $E$ remain f\/ixed we can admit  values of integers  $m$
and $n$ such that $pn+qm$ is a constant. One possibility that suggests itself is
that $n\rightarrow n+q$, $m\rightarrow m-p$. To see that this is achievable via
dif\/ferential operators we need only consider
\[ \Psi =L^{a_1}_n(z_1)L^{a_2}_m(z_2),\qquad  z_1=\mu _1x^2,\qquad z_2=\mu _2y^2.\]

We now note the recurrence formulas for Laguerre polynomials viz.
\[ x\frac{d}{ dx}L^\alpha _p(x)=pL^\alpha _p(x)-(p+\alpha )L^\alpha _{p-1}(x)=(p+1)L^\alpha _{p+1}(x)-(p+1+\alpha -x)L^\alpha _p(x).\]
Because of the separation equations we can associate $p$ with a dif\/ferential
operator for both $m$ and $n$ in the expression for $\Psi$. We do not change
the coef\/f\/icients of $L^\alpha _{p\pm 1}(x)$. Therefore we can raise or lower the
indices $m$ and $n$ in $\Psi $ using dif\/ferential operators. In particular we can
perform the transformation  $n\rightarrow n+q$, $m\rightarrow m-p$
and preserve the energy eigenvalue. To see how this works we observe the
formulas
\begin{gather*}
 D^+(\mu _1,x)X_n=\left(\partial ^2_x-2x\mu _1\partial _x-\mu _1+\mu ^2_1x^2 +\frac{\frac14
-\alpha_1^2}{ x^2}\right)X_n=-4\mu _1(n+1)X_{n+1},\\
 D^-(\mu _2,y)Y_m=\left(\partial ^2_y+2y\mu _2\partial _y+\mu _2+\mu ^2_2y^2+ \frac{\frac14
-\alpha_2^2}{ y^2}\right)Y_m=-4\mu _2(m+\alpha_2)Y_{m-1}.
\end{gather*}
In particular, if we make the choice $\mu _1=2\mu$, $\mu _2=\mu $
then the operator $D^+(2\mu ,x)D^-(\mu,y)^2$ transforms $X_nY_m$ to
$-128\mu^3(n+1)(m+\alpha_2)(m-1+\alpha_2)X_{n+1}Y_{m-2}$. We see that this preserves the energy
eigenspace. Thus can easily be  extended  to the case when $\mu _1=p\mu $
and $\mu _2=q\mu $ for $p$ and $q$ integers. A suitable operator is
$D^+(p\mu ,x)^qD^-(q\mu ,y)^p$. Hence we have constructed a dif\/ferential operator that commutes with $H$! The
caged oscillator is quantum superintegrable. This works to prove superintegrability in all dimensions as
we need only take coordinates pairwise.

\medskip

\noindent
{\bf Remark.} This second proof of superintegrability, using dif\/ferential recurrence relations for Laguerre polynomials is much more transparent than the canonical operator proof, and it generalizes immediately to all dimensions. Unfortunately, this oscillator system is very  simple and the recurrence relation approach is more dif\/f\/icult to implement for more complicated potentials. Special function recurrence relations have to be worked out. Also, we only verif\/ied explicitly the commutivity on an eigenbasis for a bound state.

\subsection[A proof of superintegrability for a deformed Kepler-Coulomb system]{A proof of superintegrability for a deformed Kepler--Coulomb system}

In \cite{PW2010} there is introduced a new family of Hamiltonians with a deformed Kepler--Coulomb potential dependent on an indexing parameter k which is shown to be related to the TTW oscillator system system via coupling constant metamorphosis. The authors showed that this system is classically superintegrable for all rational $k$. Here we demonstrate that this system is also quantum superintegrable. The proof follows easily from the canonical equations for the system.

The quantum TTW system is $H\Psi=E\Psi$ with $H$  given by (\ref{TIS1}) where
\begin{gather}\label{TTWa} u_1=R,\qquad u_2=\theta,\qquad f_1=e^{2R},\qquad f_2=0,\qquad v_1=\alpha e^{4R},\\
 v_2=\frac{\beta}{\cos^2(k\theta)}+\frac{\gamma}{\sin^2(k\theta)}=\frac{2(\gamma+\beta)}{\sin^2(2k\theta)}
 +\frac{2(\gamma-\beta)\cos(2k\theta)}{\sin^2(2k\theta)}.\nonumber
 \end{gather}
(Setting $r=e^R$ we get the usual expression for this system in polar coordinates $r$, $\theta$.) In our paper \cite{KKM10a} we used the canonical form for symmetry operators to establish the superintegrability of this system.  Our procedure was,
based on the results of \cite{KMPog10} for the classical case, to postulate expansions of  $F$, $G$ in f\/inite series
\begin{gather*}
F=\sum_{a,b,c} A_{a,b,c}E_{a,b,c}(R,\theta),\qquad G=\sum_{a,b,c} B_{a,b,c}E_{a,b,c}(R,\theta),\\
  E_{a,b,0}=e^{2aR}\sin^b(2k\theta),\qquad E_{a,b,1}=e^{2aR}\sin^b(2k\theta)\cos(2k\theta).\nonumber
\end{gather*}
The sum is taken over terms of the form $a=a_0+m$, $b=b_0+n$, and $c=0,1$, where $m$, $n$ are integers,  $a_0$ is a positive integer and $b_0$ is a negative integer. Here $F$, $G$ are the solutions of the canonical form equations (\ref{ABC}), (\ref{eqn1}), (\ref{eqn2}) for the TTW system. We succeeded in f\/inding  f\/inite series solutions for all rational~$k$, and this proved superintegrability.

In \cite{PW2010} the authors point out that under the St\"ackel transformation determined by the potential $U=e^{2R}$ the TTW system transforms to the equivalent system ${\hat H}\Psi=-\alpha\Psi$, where
\begin{gather}\label{TTS2} {\hat H}=
\frac{1}{e^{4R}}\left(\partial^2_{R}+\partial^2_{\theta}-E\exp(2R)
+\frac{\beta}{\cos^2(k\theta)}+\frac{\gamma}{\sin^2(k\theta)}\right).
\end{gather}
Then, setting $r=e^{2R},\ \phi=2\theta$, we f\/ind the deformed Kepler--Coulomb system.
\[ \left(\partial^2_r+\frac{1}{r}+\frac{1}{r^2}\partial^2_\phi-\frac{E}{4r}
+\frac{\beta}{4r^2\cos^2(k\phi/2)}+\frac{\gamma}{4r^2\sin^2(k\phi/2)}\right)\Psi=-\frac{\alpha}{4}\Psi.\]

From the canonical equations, it is virtually immediate that system (\ref{TTS2}) is superintegrable. Indeed the only dif\/ference between the
functions (\ref{TTWa}) def\/ining the TTW system and the functions def\/ining the St\"ackel transformed system is that $f_1$
and $v_1$ are replaced by  $f_1=\exp(4R)$ and $v_1=-E \exp(2R)$ (we can set $E=H$) and the former $H$ is replaced by $-\alpha$. From this we f\/ind that  the only dif\/ference
between the canonical equations for the TTW system and the canonical equations for the transformed system is that $\alpha$ becomes~$-{\hat H}$ and~$H$ becomes~$-\alpha$.
Since our proof of the superintegrability of the TTW system  did not
depend on the values of $H $ and $\alpha$, the same argument
shows the superintegrability of the transformed system.  Also, any solution
(i.e., a second commuting operator) for the TTW system
gives  rise to a corresponding operator for the transformed system  by replacing~$\alpha$ and~$H$ with $-{\hat H}$ and $-\alpha$, respectively.

\medskip
\noindent
{\bf Example.}
For the deformed Kepler system with $k=2$, we take
$p=2$ and $q=1$ and use Gaussian elimination to f\/ind a solution with nonzero coef\/f\/icients
\[
A_{-2, 1, 0}, \ A_{-1, 1, 0}, \
B_{-2, 0, 0}, \ B_{-2, 0, 1}, \ B_{-1, 0, 0}, \ B_{-1, 0, 1}, \ B_{0, 0, 1},
\]
in which $A_{-2,1,0}$ and $A_{-1,1,0}$ are two independent parameters.  To achieve
the lowest order symmetry and ensure that the $A$, $B$, $C$ and $D$ are polynomial
in ${\hat L}_2$ and $\hat H$, we choose $A_{-2,1,0}=32({\hat L}_2-4)$ and $A_{-1,1,0}=0$ and f\/ind,
\begin{gather*}
A = 32e^{-4R}\sin(4\theta){\hat L}_2-128e^{-4R}\sin(4\theta),\\
B = 8e^{-4R}\cos(4\theta){\hat L}_2^2
 + (8(\beta-\gamma)e^{-4R}+4Ee^{-2R}\cos(4\theta)-8e^{-4R}\cos(4\theta)){\hat L}_2 \\
\phantom{B=}{}
 - 96e^{-4R}\cos(4\theta)-16Ee^{-2R}\cos(4\theta)+40(\gamma-\beta)e^{-4R}+4E(\beta-\gamma)e^{-2R},  \\
C = -8e^{-4R}\sin(4\theta){\hat L}_2^2-4\sin(4\theta){\hat H}{\hat L}_2
 + 8(e^{-4R}-Ee^{-2R})\sin(4\theta){\hat L}_2 +20\sin(4\theta){\hat H} \\
\phantom{C=}{}  +96e^{-4R}\sin(4\theta) +32Ee^{-2R}\sin(4\theta) -E^2\sin(4\theta), \\
D = -32e^{-4R}\cos(4\theta)L_2^2 - 8 \cos(4\theta){\hat H}{\hat L}_2 \\
\phantom{D=}{}
 + (128e^{-4R}\cos(4\theta)-12Ee^{-2R}\cos(4\theta)+16(\gamma-\beta)e^{-4R}){\hat L}_2
 + 24\cos(4\theta){\hat H} \\
\phantom{D=}{}
 + 48(\beta-\gamma)e^{-4R} + 48Ee^{-2R}\cos(4\theta)-4E(\gamma-\beta)e^{-2R} + 2E^2\cos(4\theta).
\end{gather*}

It has been explicitly checked, using Maple, that the f\/ifth order operator obtained from these expressions
commutes with $\hat H$.

\subsection{A proof of superintegrability for the caged oscillator on the hyperboloid}
Using this same idea we can f\/ind a new superintegrable system on the 2-sheet hyperboloid by taking a St\"ackel transformation of the quantum caged oscillator $H\Psi=E\Psi$,
(\ref{cagedoscillator}) in Cartesian coordinates $u_1$, $u_2$  by multiplying with the potential $U=1/u_1^2$. The transformed system is
\[  u_1^2\left(\partial_1^2+\partial_2^2+\omega^2u_1^2(p^2u_1^2+q^2u_2^2)-Eu_1^2+\alpha_2\frac{u_1^2}{u_2^2}\right)\Psi=-\alpha_1\Psi.\]
We embed  this system as the upper sheet $s_0>0$ of the 2-sheet hyperboloid $s_0^2-s_1^2-s_2^2=1$ in 3-dimensional Minkowski space with  Minkowski metric  $ds^2=-ds_0^2+ds_1^2+ds_2^2$,
via the coordinate transformation
\[  s_0=\frac{1+u_1^2+u_2^2}{2u_1},\qquad s_1=\frac{1-u_1^2-u_2^2}{2u_1},\qquad s_2=\frac{u_2}{u_1},\qquad u_1>0.\]
Then the potential for the transformed system is
\begin{gather}\label{quantumhyperboloid} {\tilde V}=\frac{\alpha_2}{s_2^2}-\frac{E}{(s_0+s_1)^2}
+\frac{\omega^2(p^2-q^2)}{(s_0+s_1)^4}+\frac{\omega^2q^2(s_0-s_1)}{(s_0+s_1)^3}.\end{gather}
This is an extension of the complex sphere system [S2], distinct from the system~(\ref{hamS2}) that we proved classically superintegrable. It is an easy consequence of the results  of~\cite{KMP10} that the classical version of  system (\ref{quantumhyperboloid}) is also superintegrable for all rationally related~$p$,~$q$. However, quantum superintegrability isn't obvious. However, from the results of Section~\ref{cagedosc}    it is virtually immediate that this new system is quantum superintegrable for all relatively prime positive integers~$p$, $q$.  This follows from  writing down the canonical equations  (\ref{eqn1}), (\ref{eqn2}), f\/irst for the    caged oscillator where
\[   f_1=1,\qquad f_2=0,\qquad v_1=\omega^2 p^2u_1^2+\frac{\alpha_1}{u_1^2},\qquad v_2=\omega^2 q^2u_2^2+\frac{\alpha_2}{u_2^2},\]
and then for the St\"ackel transformed system where
\[  f_1=\frac{1}{u_1^2},\qquad f_2=0,\qquad v_1=\omega^2 p^2u_1^2-E,\qquad v_2=\omega^2 q^2u_2^2+\frac{\alpha_2}{u_2^2}.\]
The equations are identical except for the switches  $\alpha_1\to -E$, $H\to -\alpha_1$. Thus our proof of quantum superintegrability for the caged oscillator carries over to show that the system on the hyperboloid is also  quantum superintegrable.

\section{Discussion} A basic issue in discovering and verifying higher order superintegrabily of classical and quantum systems is the dif\/f\/iculty of manipulating high order constants of the motion and, particularly, higher order partial dif\/ferential operators. We have described several approaches to simplify such calculations. Although our primary emphasis  in this paper was  to develop tools for verifying classical and quantum superintegrabity at all orders, we have presented many new results. The classical Eucidean systems [E8], [E17] and  most of  the examples of superintegrability for spaces with non-zero scalar curvature are new. We have explored the limits of the construction of classical superintegrable systems via the methods of Section \ref{subsection1} for the case $n=2$. Further use of this method will require looking at $n>2$, where new types of behavior occur, such as appearance of superintegrable systems that are not conformally f\/lat.  We also developed the use of the canonical form for a symmetry operator to prove quantum superintegrability. We applied the method to the $n=2$ caged anisotropic oscillator to give the f\/irst proof of  quantum superintegrability for all rational~$k$. We used the St\"ackel transform together with the canonical equations to give the f\/irst proofs of quantum superintegrability for all rational~$k$ of a 2D deformed Kepler--Coulomb system, and of the caged anisotropic oscillator on the 2-hyperboloid. Then we introduced a new approach to proving quantum superintegrability via recurrence relations obeyed by the energy eigenfunctions of a quantum system, and gave an alternate proof of the superintegrabilty for all rational~$k$ of the caged anisotropic oscillator. This proof clearly extends to all $n$. The recurrences for the caged oscillator are particularly simple, but the method we presented shows great promise for broader application. Clearly, we are just at the beginning of the process of discovery and classif\/ication of higher order superintegrable systems in all dimensions~$n>2$.

\pdfbookmark[1]{References}{ref}
\LastPageEnding

\end{document}